\documentclass[aps,prd,twocolumn,superscriptaddress,showpacs,showkeys,floatfix]{revtex4}

\usepackage[dvips]{graphicx}
\usepackage{xspace}
\usepackage{latexsym}
\usepackage{amssymb}
\usepackage{times}
\usepackage{mathptm}
 
\graphicspath{{./figs/}}

\begin{document}

\newcommand{\superk}    {Super-Kamiokande\xspace}       
\newcommand{\nue}       {$\nu_{e}$\xspace}
\newcommand{\numu}      {$\nu_{\mu}$\xspace}
\newcommand{\nutau}     {$\nu_{\tau}$\xspace}
\newcommand{\nusterile} {$\nu_{sterile}$\xspace}
\newcommand{\mutau}     {$\nu_\mu \rightarrow \nu_{\tau}$\xspace}
\newcommand{\musterile} {$\nu_\mu \rightarrow \nu_{sterile}$\xspace}
\newcommand{\dms}       {$\Delta m^2$\xspace}
\newcommand{\sstt}      {$\sin^2 2 \theta$\xspace}
\newcommand{\asymerr}[2]{\ooalign{{\scriptsize \raisebox{4pt}{+~#1}}\crcr
{\scriptsize \raisebox{-4pt}{--~#2}}}}

\title{Search for Matter-Dependent Atmospheric Neutrino Oscillations in Super-Kamiokande}

\date{\today}

\newcounter{foots}
\newcounter{notes}
\newcommand{\authoraticrr}{$^{1}$}
\newcommand{\authoratncen}{$^{2}$}
\newcommand{\authoratbu}{$^{3}$}
\newcommand{\authoratbupenn}{$^{3,\dagger}$}
\newcommand{\authoratbnl}{$^{4}$}
\newcommand{\authoratuci}{$^{5}$}
\newcommand{\authoratcsu}{$^{6}$}
\newcommand{\authoratcnu}{$^{7}$}
\newcommand{\authoratduke}{$^{8}$}
\newcommand{\authoratgmu}{$^{9}$}
\newcommand{\authoratgifu}{$^{10}$}
\newcommand{\authoratuh}{$^{11}$}
\newcommand{\authoratui}{$^{12}$}
\newcommand{\authoratkek}{$^{13}$}
\newcommand{\authoratkekicrr}{$^{13,1}$}
\newcommand{\authoratkekkashiwa}{$^{13,\ddagger}$}
\newcommand{\authoratkobe}{$^{14}$}
\newcommand{\authoratkyoto}{$^{15}$}
\newcommand{\authoratkyototriumf}{$^{15,\S}$}
\newcommand{\authoratlanluci}{$^{16,5}$}
\newcommand{\authoratlsu}{$^{17}$}
\newcommand{\authoratumd}{$^{18}$}
\newcommand{\authoratduluth}{$^{19}$}
\newcommand{\authoratmiyagi}{$^{20}$}
\newcommand{\authoratnagoya}{$^{21}$}
\newcommand{\authoratsuny}{$^{22}$}
\newcommand{\authoratniigata}{$^{23}$}
\newcommand{\authoratokayama}{$^{24}$}
\newcommand{\authoratosaka}{$^{25}$}
\newcommand{\authoratseoul}{$^{26}$}
\newcommand{\authoratshizuoka}{$^{27}$}
\newcommand{\authoratshizuokaseika}{$^{28}$}
\newcommand{\authoratskku}{$^{29}$}
\newcommand{\authorattohoku}{$^{30}$}
\newcommand{\authorattokai}{$^{31}$}
\newcommand{\authorattit}{$^{32}$}
\newcommand{\authorattokyo}{$^{33}$}
\newcommand{\authorattsinghua}{$^{34}$}
\newcommand{\authoratwarsaw}{$^{35}$}
\newcommand{\authoratwarsawuci}{$^{35,5}$}
\newcommand{\authoratuw}{$^{36}$}
\newcommand{\authoratuwduluth}{$^{36,19}$}
\newcommand{\addressoficrr}[1]{$^{1}$ #1 }
\newcommand{\addressofncen}[1]{$^{2}$ #1 }
\newcommand{\addressofbu}[1]{$^{3}$ #1 }
\newcommand{\addressofbnl}[1]{$^{4}$ #1 }
\newcommand{\addressofuci}[1]{$^{5}$ #1 }
\newcommand{\addressofcsu}[1]{$^{6}$ #1 }
\newcommand{\addressofcnu}[1]{$^{7}$ #1 }
\newcommand{\addressofduke}[1]{$^{8}$ #1 }
\newcommand{\addressofgmu}[1]{$^{9}$ #1 }
\newcommand{\addressofgifu}[1]{$^{10}$ #1 }
\newcommand{\addressofuh}[1]{$^{11}$ #1 }
\newcommand{\addressofui}[1]{$^{12}$ #1 }
\newcommand{\addressofkek}[1]{$^{13}$ #1 }
\newcommand{\addressofkobe}[1]{$^{14}$ #1 }
\newcommand{\addressofkyoto}[1]{$^{15}$ #1 }
\newcommand{\addressoflanl}[1]{$^{16}$ #1 }
\newcommand{\addressoflsu}[1]{$^{17}$ #1 }
\newcommand{\addressofumd}[1]{$^{18}$ #1 }
\newcommand{\addressofduluth}[1]{$^{19}$ #1 }
\newcommand{\addressofmiyagi}[1]{$^{20}$ #1 }
\newcommand{\addressofnagoya}[1]{$^{21}$ #1 }
\newcommand{\addressofsuny}[1]{$^{22}$ #1 }
\newcommand{\addressofniigata}[1]{$^{23}$ #1 }
\newcommand{\addressofokayama}[1]{$^{24}$ #1 }
\newcommand{\addressofosaka}[1]{$^{25}$ #1 }
\newcommand{\addressofseoul}[1]{$^{26}$ #1 }
\newcommand{\addressofshizuoka}[1]{$^{27}$ #1 }
\newcommand{\addressofshizuokaseika}[1]{$^{28}$ #1 }
\newcommand{\addressofskku}[1]{$^{29}$ #1 }
\newcommand{\addressoftohoku}[1]{$^{30}$ #1 }
\newcommand{\addressoftokai}[1]{$^{31}$ #1 }
\newcommand{\addressoftit}[1]{$^{32}$ #1 }
\newcommand{\addressoftokyo}[1]{$^{33}$ #1 }
\newcommand{\addressoftsinghua}[1]{$^{34}$ #1 }
\newcommand{\addressofwarsaw}[1]{$^{35}$ #1 }
\newcommand{\addressofuw}[1]{$^{36}$ #1 }

\def\pennnow{$\dagger$}
\def\kashiwanow{$\ddagger$}
\def\triumfnow{\S}
\author{
{\bf The Super-Kamiokande Collaboration} \\
\vspace{0.2cm}
K.~Abe\authoraticrr,
Y.~Hayato\authoraticrr,
T.~Iida\authoraticrr,
M.~Ikeda\authoratokayama,
J.~Kameda\authoraticrr,
Y.~Koshio\authoraticrr,
A.~Minamino\authoraticrr,
M.~Miura\authoraticrr,
S.~Moriyama\authoraticrr,
M.~Nakahata\authoraticrr
% ...
\\
\smallskip
\smallskip
\footnotesize
\it
\addressoficrr{Kamioka Observatory, Institute for Cosmic Ray Research, 
University of Tokyo, Kamioka, Gifu, 506-1205, Japan}\\
\addressofncen{Research Center for Cosmic Neutrinos, Institute for Cosmic 
Ray Research, University of Tokyo, Kashiwa, Chiba 277-8582, Japan}\\
\addressofbu{Department of Physics, Boston University, Boston, MA 02215, 
USA}\\
\addressofbnl{Physics Department, Brookhaven National Laboratory, Upton, 
NY 11973, USA}\\
\addressofuci{Department of Physics and Astronomy, University of 
California, Irvine, Irvine, CA 92697-4575, USA }\\
\addressofcsu{Department of Physics, California State University, 
Dominguez Hills, Carson, CA 90747, USA}\\
\addressofcnu{Department of Physics, Chonnam National University, Kwangju 
500-757, Korea}\\
\addressofduke{Department of Physics, Duke University, Durham, NC 27708, 
USA} \\
\addressofgmu{Department of Physics, George Mason University, Fairfax, VA 
22030, USA }\\
\addressofgifu{Department of Physics, Gifu University, Gifu, Gifu 
501-1193, Japan}\\
\addressofuh{Department of Physics and Astronomy, University of Hawaii, 
Honolulu, HI 96822, USA}\\
\addressofui{Department of Physics, Indiana University, Bloomington,
  IN 47405-7105, USA} \\
\addressofkek{High Energy Accelerator Research Organization (KEK), 
Tsukuba, Ibaraki 305-0801, Japan }\\
\addressofkobe{Department of Physics, Kobe University, Kobe, Hyogo 
657-8501, Japan}\\
\addressofkyoto{Department of Physics, Kyoto University, Kyoto 606-8502, 
Japan}\\
\addressoflanl{Physics Division, P-23, Los Alamos National Laboratory, Los 
Alamos, NM 87544, USA }\\
\addressoflsu{Department of Physics and Astronomy, Louisiana State 
University, Baton Rouge, LA 70803, USA }\\
\addressofumd{Department of Physics, University of Maryland, College Park, 
MD 20742, USA }\\
\addressofduluth{Department of Physics, University of Minnesota, Duluth, 
MN 55812-2496, USA}\\
\addressofmiyagi{Department of Physics, Miyagi University of Education, 
Sendai, Miyagi 980-0845, Japan}\\
\addressofnagoya{Solar Terrestrial Environment Laboratory, Nagoya University, Nagoya, Aichi 
464-8602, Japan}\\
\addressofsuny{Department of Physics and Astronomy, State University of 
New York, Stony Brook, NY 11794-3800, USA}\\
\addressofniigata{Department of Physics, Niigata University, Niigata, 
Niigata 950-2181, Japan }\\
\addressofokayama{Department of Physics, Okayama University, Okayama, 
Okayama 700-8530, Japan} \\
\addressofosaka{Department of Physics, Osaka University, Toyonaka, Osaka 
560-0043, Japan}\\
\addressofseoul{Department of Physics, Seoul National University, Seoul 
151-742, Korea}\\
\addressofshizuoka{Department of Systems Engineering, Shizuoka University, 
Hamamatsu, Shizuoka 432-8561, Japan}\\
\addressofshizuokaseika{Department of Informatics in Social Welfare, Shizuoka University 
of Welfare, Yaizu, Shizuoka, 425-8611, Japan}\\
\addressofskku{Department of Physics, Sungkyunkwan University, Suwon 
440-746, Korea}\\
\addressoftohoku{Research Center for Neutrino Science, Tohoku University, 
Sendai, Miyagi 980-8578, Japan}\\
\addressoftokai{Department of Physics, Tokai University, Hiratsuka, 
Kanagawa 259-1292, Japan}\\
\addressoftit{Department of Physics, Tokyo Institute for Technology, 
Meguro, Tokyo 152-8551, Japan }\\
\addressoftokyo{The University of Tokyo, Tokyo 113-0033, Japan }\\
\addressoftsinghua{Department of Engineering Physics, Tsinghua University, Beijing, 100084, China}\\
\addressofwarsaw{Institute of Experimental Physics, Warsaw University, 
00-681 Warsaw, Poland }\\
\addressofuw{Department of Physics, University of Washington, Seattle, WA 
98195-1560, USA}
\\
{\pennnow}{Present address: Center for Gravitational Wave Physics, 
Pennsylvania State 
University, University Park, PA 16802, USA} \\
}



\begin{abstract}
  
  We consider \mutau oscillations in the context of the Mass Varying Neutrino (MaVaN) 
model, where the neutrino mass can vary depending on the 
electron density along the flight path of the neutrino. Our analysis assumes a 
mechanism with dependence only upon the electron density, hence ordinary matter 
density, of the medium through which the neutrino travels. Fully-contained, 
partially-contained and upward-going muon atmospheric neutrino data from the 
Super--Kamiokande detector, taken from the entire SK--I period of 1489 live days, are 
compared to MaVaN model predictions.  We find that, for the case of 2-flavor 
oscillations, and for the specific models tested, oscillation independent of electron 
density is favored over density dependence. Assuming maximal mixing, the best-fit case 
and the density-independent case do not differ significantly.

\end{abstract}


\maketitle

\section{Introduction}
\label{sec:introduction}
Neutrino oscillations result when at least one neutrino has mass different from the others, so that neutrinos mass eigenstates are distinct from the flavor eigenstates.  For atmospheric neutrinos, \mutau 
oscillations~\cite{sk:discovery98} are strongly favored over $\nu_\mu\rightarrow\nu_s$ oscillations~\cite{Fukuda:2000np} and other exotic mechanisms for neutrino disappearance~\cite{Ashie:2004mr}.  Neutrino oscillation without the introduction of sterile flavors has also been sufficient to resolve the solar neutrino problem~\cite{Fukuda:2001nk,Ahmad:2002jz}.  

Super-Kamiokande previously reported best fit parameters for atmospheric two-flavor 
\mutau oscillations, providing an explanation of the atmospheric neutrino 
anomaly~\cite{Ashie:2005ik}. In Super-Kamiokande, where previous analyses always 
considered only geometric path length independent of medium, neutrino disappearance 
effects are strikingly evident for upward going neutrinos, which have passed through 
many km of rock, in comparison with downward-going neutrinos, which travel mainly 
through air. We note that other experiments that observed neutrino oscillations, such as KAMLAND~\cite{Araki:2004mb}, K2K~\cite{Ahn:2006zz}, and MINOS~\cite{Michael:2006rx}, detected neutrinos whose path was almost entirely through rock.

Here we consider a possible consequence of the Mass Varying Neutrino (MaVaN) 
model~\cite{Kaplan:2004dq}.  In this model, the neutrino mass can vary depending on 
the matter density along the path of the neutrino.  MaVaNs could provide a source for 
the dark energy~\cite{Zurek:2004vd}.  Here we will assume that the mass variation of 
the neutrinos depends only upon the electron density of the environment, a possible 
side effect from radiative couplings of active neutrinos and 
electrons~\cite{Kaplan:2004dq}. (In ordinary matter comprising the Earth's atmosphere 
and interior, overall matter density and baryon density, considered in other MaVaN 
scenarios, are essentially proportional to electron density.) If this hypothesis is 
accurate, it may be possible to probe the mass dependence with Super--Kamiokande data.

We used atmospheric neutrino data from the Super-Kamiokande-I running period (1996--2001) to test whether \mutau oscillations have an apparent dependence on the electron density of the material the neutrino passes through.  For this analysis, we assumed that the mass squared difference is proportional to some power of the electron density, $\Delta m^2_{eff}\sim\rho^n$ .  We also assumed that the mixing angle is constant for all media.

\section{Data Analysis}
\label{sec:dataanalysis}
Super-Kamiokande is a water Cherenkov experiment located within 
Mt. Ikeno-yama in central Japan, under 2700 meters water equivalent rock overburden.  It has a cylindrical design, holds 50 kilotons of water, and is divided into two optically separated sections by a structural framework that supports the photomultiplier tubes (PMTs). During the SK-I running period from which the data analyzed here came, the detector had an inner detector (ID) equipped with 11146 50\-cm PMTs aimed inward, and an outer detector (OD) volume instrumented with 1885 20\-cm PMTs, aimed outward and equipped with wavelength-shifting plastic plates. The OD functions primarily as a veto counter, tagging charged particles that enter or exit the ID.  Within the ID we define a central 22.5 kiloton fiducial volume, within which detector response is expected to be uniform. Fully-contained (FC) neutrino events are those where interaction products are observed in the ID, with no significant correlated activity in the OD, while partially-contained (PC) events are those where some interaction products exit the ID. Upward-going muon events are those where a penetrating particle travelling in the upward direction enters and either stops or passes through the detector, and are attributed to muons produced by neutrino interactions in the surrounding rock. In general terms, FC, PC, and upward muon events represent successively higher energy samples of neutrino interactions, ranging from 200 MeV for the lowest energy FC events to above 1 TeV for the highest energy upward-going muons. Further details regarding the Super-Kamiokande detector design, operation, calibrations, and data reduction can be found in~\cite{Fukuda:2002uc,Ashie:2005ik}.

Super-Kamiokande-I (SK-I) data taking for physics analysis began on May 17, 1996 and lasted until a planned shutdown for refurbishing on July 16, 2001, including a total of 1489.2 days of effective livetime (92 kiloton-years) for FC and PC events, and 1645.9 days for upward-going muon events.  The full Monte Carlo (MC) event sample generated for comparison is equivalent to a 100 live-year period.  The SK-I database used in the analyses presented here has the following statistics: for FC events, 12180 data and 13676.7 MC events in a livetime-scaled sample, for PC events, 911 data and 1129.6 MC events. The upward-going muon statistics are: for stopping muon tracks (those that enter but do not exit the detector), 417.7 data and 713.5 MC events, and for through-going muon tracks, 1841.6 data and 1669.5 MC events, after background subtraction for near-horizontal downward-going muons as described in 
ref.~\cite{Ashie:2005ik}. 

Two different analyses are considered here. The first analysis does not take into account path length in air for neutrino oscillations.  Instead, it assumes MaVan-type oscillations only occur in high-density matter, taking into account surface, crust, mantle, and Earth core densities. In other words, the neutrino flight path length is in effect taken to be only that portion of the geometric path that lies in high density matter. Because the path length systematic error is not relevant in this test, it is removed from the list of 39 systematic errors that was used in previous Super-Kamiokande analyses. The second analysis also takes into account path length in air and its density in considering neutrino oscillations effects in the context of the MaVaN model.

A detailed description of the SK-I atmospheric neutrino chi-squared zenith angle analysis using the ``pull" method~\cite{fogli-pull} can be found in~\cite{Ashie:2005ik}, and the same methods are applied here.  A profile of the mountain surrounding Super-Kamiokande was taken from topological maps to determine the downward neutrino path length in rock.  The maps used are the United States Geological Survey agency's Digital Elevation Maps~\cite{usgs}, with data points at 7.5 minute (about 30 meters) spacing.

In the conventional 2-flavor \mutau oscillation framework, the oscillation probability can be written as
$$P(\nu_\mu\rightarrow\nu_\tau)=\sin^2{2\theta}\sin^2\left(\frac{{1.27\Delta m^2 L}}{E_\nu}\frac{{\rm GeV}}{{\rm km}\cdot{\rm eV}^2}\right),$$
where $L$ is the distance travelled by the neutrino between production and detection, and $E_\nu$ is the neutrino energy, while the oscillation parameters are the mixing angle $\theta$, and the mass difference squared, $\Delta m^2=|m^2_{2}-m^2_{3}|$ in the usual nomenclature.
To test for density dependence, we replaced $\Delta m^2$ in the equation above with an effective mass difference that is proportional to the electron density of the medium, $\Delta m^2\rightarrow\Delta m^2\times \left(\frac{\rho_e}{\rho_o}\right)^n$, where $\rho_e$ is the electron density of the matter in neutrino trajectory, and $n$ parameterizes the density dependence. For neutrinos passing through layers of matter with different density, the path length in each layer is taken into account. $\rho_0$ is set at $6.02\times10^{23}{\rm{ e}}/{\rm{cm}}^3$.  Several different values of $n$ are tested here, as described below.  
To approximate the mass density of the Earth for upward traveling neutrinos, the Preliminary Earth Reference Model (PREM)~\cite{Dziewonski:1981xy} is used.  The electron density is taken to be the mass density multiplied by the charge-to-mass ratio~\cite{Bahcall:1997jc}.  

Atmospheric neutrinos are produced at altitudes around 15 km above sea level. A neutrino travelling nearly vertically downward thus passes through about 15 km of air (low-density matter) followed by 1-2 km of rock (high density matter) before interacting within the detector, while a nearly-horizontal downward-going neutrino travels through over 200 km of air and about 10 km of rock.  Thus, a downward neutrino has 8-20 times as much path length in air as in rock.  If oscillation only occurs in high density matter, the effective path length will be shorter than the geometric path length travelled by the neutrino from point of production to detection. In that case, the standard oscillation model, which uses the full geometric path length, assumes more oscillation cycles occur for a given $\Delta m^2$, and oscillation effects are more significant for neutrinos at low energies. Analyses both including, and neglecting, the air path length are considered here.

\subsection{Analysis Neglecting Air path length}
\label{ssec:noairanalysis}

Density dependent oscillation models were first tested neglecting the air path length 
and using specified values of $n$ suggested by theorists~\cite{pcomm1}.  The results 
are shown in Table~\ref{tbl:rockonlychicomp} for two cases: fits constrained to lie in 
the physical region, where maximal mixing is assumed for all densities, and with this 
constraint relaxed, where mixing angle is allowed to vary.  All models tested are 
ruled out, with confidence level better than 99.9\%, when compared to fits assuming 
oscillations independent of density of the medium (including air). 

\begin{table*}[htb] 
\centering
 \caption[Comparison of $\chi^2_{min}$ for different MaVaN models with no air path 
length.]{Comparisons of $\chi^2$ Values for different models without air path.  There 
are a total of 178 degrees of freedom in the $\chi^2$ (see ref.~\cite{Ashie:2005ik}).  
Note that $n=0$ is {\em not} equivalent to oscillations independent of medium, because 
here we do not allow for any oscillations in air.}

\begin{tabular}{|c|c|c|c|c|c|c|c|}
\hline
Model&\multicolumn{4}{c|}{Physical Region}&\multicolumn{3}{c|}{Unphysical Region}\\
$\Delta m^2\times(\rho/\rho_o)^n$&$\chi^2_{min}$&$\Delta m^2$(eV$^2$)&$\sin^2(2\theta)$&$\Delta\sigma$&$\chi^2_{min}$&$\Delta m^2$(eV$^2$)&$\sin^2(2\theta)$\\
\hline
$n=2$&203.4&$4.84\times10^{-4}$&1.00&5.3&203.1&$4.67\times10^{-4}$&1.02\\ 
$n=1$&194.7&$1.12\times10^{-3}$&1.00&4.4&194.1&$1.19\times10^{-3}$&1.04\\ 
$n=\frac{2}{3}$&192.5&$1.46\times10^{-3}$&1.00&4.2&191.7&$1.46\times10^{-3}$&1.04\\ 
$n=\frac{1}{3}$&190.5&$1.60\times10^{-3}$&1.00&3.9&189.5&$1.60\times10^{-3}$&1.04\\ 
$n=0$&189.8&$2.66\times10^{-3}$&1.00&3.8&188.9&$2.66\times10^{-3}$&1.04\\ 
$n=-\frac{1}{3}$&188.5&$3.16\times10^{-3}$&1.00&3.7&187.5&$2.92\times10^{-3}$&1.04\\ 
$n=-\frac{2}{3}$&187.9&$3.12\times10^{-3}$&1.00&3.6&187.0&$3.12\times10^{-3}$&1.04\\ 
$n=-1$&188.5&$3.80\times10^{-3}$&1.00&3.7&187.6&$3.80\times10^{-3}$&1.04\\ 
$n=-2$&190.1&$4.73\times10^{-3}$&1.00&3.9&188.7&$4.73\times10^{-3}$&1.05\\ 
\hline
\end{tabular}
\label{tbl:rockonlychicomp}
\end{table*}

Next, we treat the density dependence as a free parameter, and find the best fit 
electron density dependence exponent $n$ for the case of matter-dependent oscillations 
assuming maximal mixing ($\sin^2(2\theta)=1$) in all densities (except air).  This 
analysis uses both $n$ and $\Delta m^2$ as the varied parameters.  $\Delta m^2$ values 
from $10^{-4}$ to $10^0$ eV$^2$ were used over 81 bins, and density dependence powers 
from $n=-3$ to $+3$ were considered, over 61 bins.  The results produce a 
$\chi^2_{min}$ = 187.3/178 d.o.f. for $(n,\Delta m^2) = (-0.30, 3.16\times10^{-3} 
{\rm eV}^2)$. Thus, the model in which air path length is neglected, {\it i.e.,} 
oscillations occur only in rock, is disfavored at about the 3.5 $\sigma$ level 
relative to standard oscillations.

\subsection{Analysis Including Air path length}

 We repeated the analysis, this time 
taking into account the path length in air. Now the $n = 0$ case, where there is no 
dependence on the electron density, corresponds to the conventional oscillations 
model, since the full geometric path length is used.

The air mass density depends greatly on the altitude of the air above sea level.  For 
simplicity, a constant value of $10^{-3}$ g/cm$^3$ (corresponding to actual air 
density at mountain altitude) was used for the mass density.

The same set of fixed $n$ values are tested again, to check various density 
dependences, this time including the air path length.  The results are listed in 
Table~\ref{tbl:rockairchicomp}.  All models tested are excluded by $\ge 3.8 
\sigma$ when compared to the density-independent case.

\begin{table*}[htb]
\begin{center}
 \caption[Comparison of $\chi^2$ for different MaVaN models including air path 
length.]{Comparisons of $\chi^2$ Values for Different Models including air path. Note 
that here, $n=0$ {\em is} equivalent to oscillations independent of medium.}
 \begin{tabular}{|c|c|c|c|c|c|c|c|}
\hline
Model&\multicolumn{4}{c|}{Physical Region}&\multicolumn{3}{c|}{Unphysical Region}\\
$\Delta m^2\times(\rho/\rho_o)^n$&$\chi^2_{min}$&$\Delta m^2$(eV$^2$)&$\sin^2(2\theta)$&$\Delta\sigma$&$\chi^2_{min}$&$\Delta m^2$(eV$^2$)&$\sin^2(2\theta)$\\
\hline
Equal medium oscillations (n=0)&175.0&$2.11\times10^{-3}$&1.00&-&174.7&$2.11\times10^{-3}$&1.02\\
$n=2$&203.4&$4.84\times10^{-4}$&1.02&5.3&203.1&$4.67\times10^{-4}$&1.00\\ 
$n=1$&194.7&$1.12\times10^{-3}$&1.00&4.4&194.1&$1.12\times10^{-3}$&1.04\\ 
$n=\frac{2}{3}$&192.4&$1.46\times10^{-3}$&1.00&4.2&191.7&$1.46\times10^{-3}$&1.04\\ 
$n=\frac{1}{3}$&189.6&$1.60\times10^{-3}$&1.00&3.8&188.7&$1.60\times10^{-3}$&1.04\\ 
$n=-\frac{1}{3}$&228.5&$7.50\times10^{-4}$&0.91&7.9&228.5&$7.50\times10^{-4}$&0.91\\ 
$n=-\frac{2}{3}$&392.1&$1.50\times10^{-4}$&0.63&14.7&392.1&$1.50\times10^{-4}$&0.70\\ 
$n=-1$&447.8&$3.63\times10^{-2}$&0.60&16.5&447.8&$3.63\times10^{-2}$&0.60\\ 
$n=-2$&447.9&$3.73\times10^{-3}$&0.60&16.5&447.9&$3.73\times10^{-3}$&0.60\\ 
\hline
\end{tabular}
\label{tbl:rockairchicomp}
\end{center}
\end{table*}

\begin{figure}[p]
\centering
\includegraphics[height=90mm]{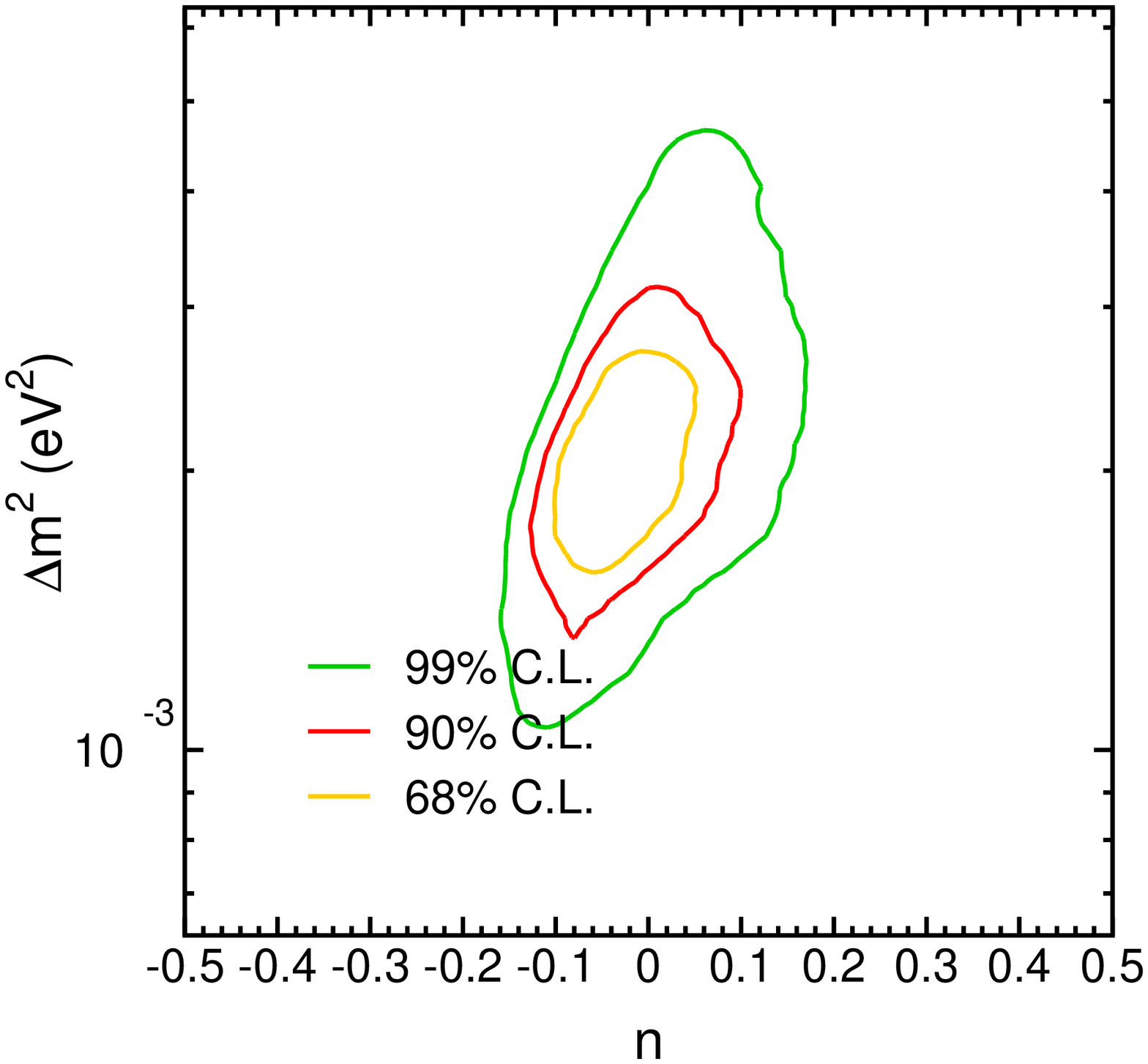}\\
\vspace{10pt}
\includegraphics[height=60mm]{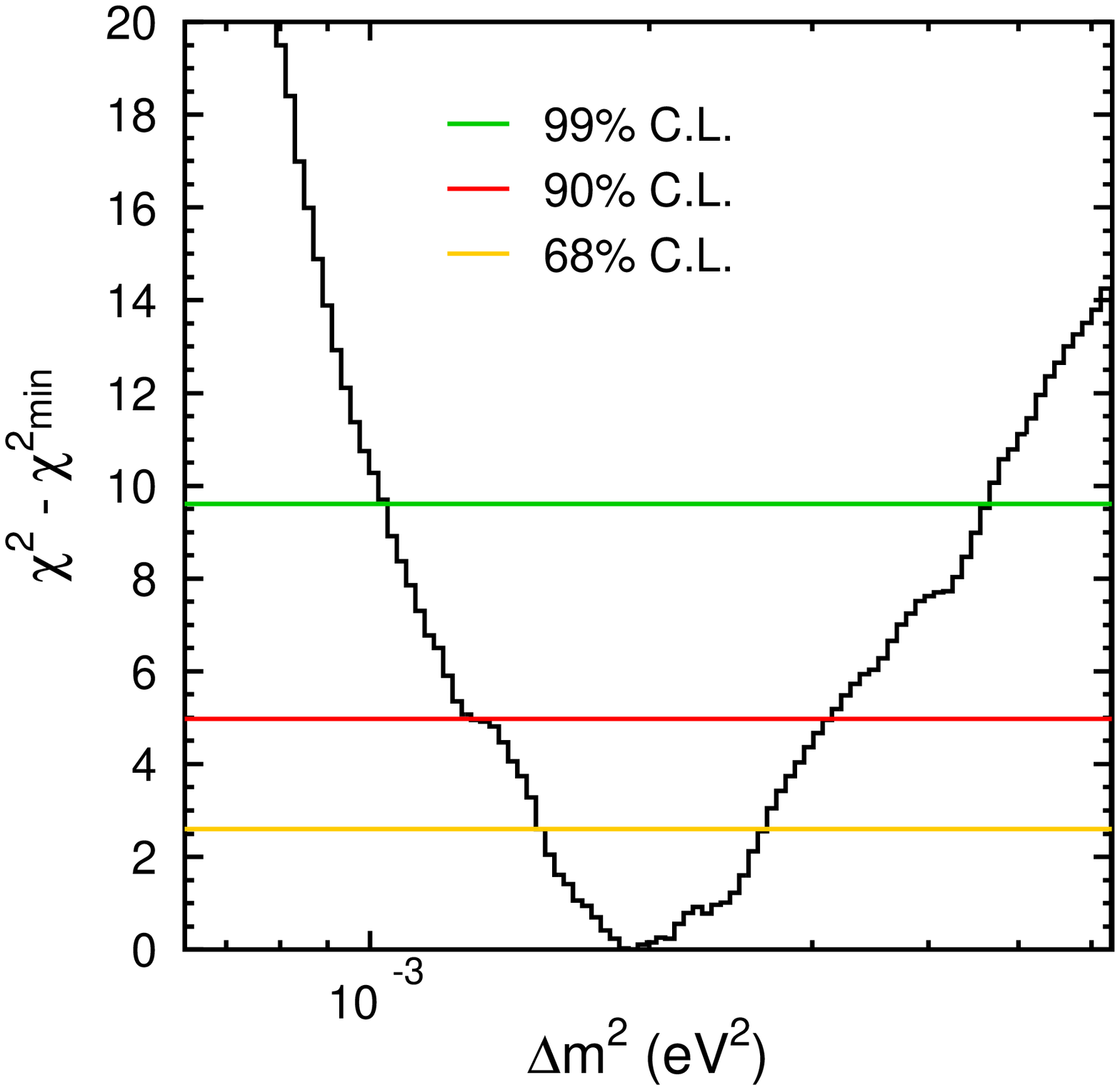}
\hspace{20pt}
\includegraphics[height=60mm]{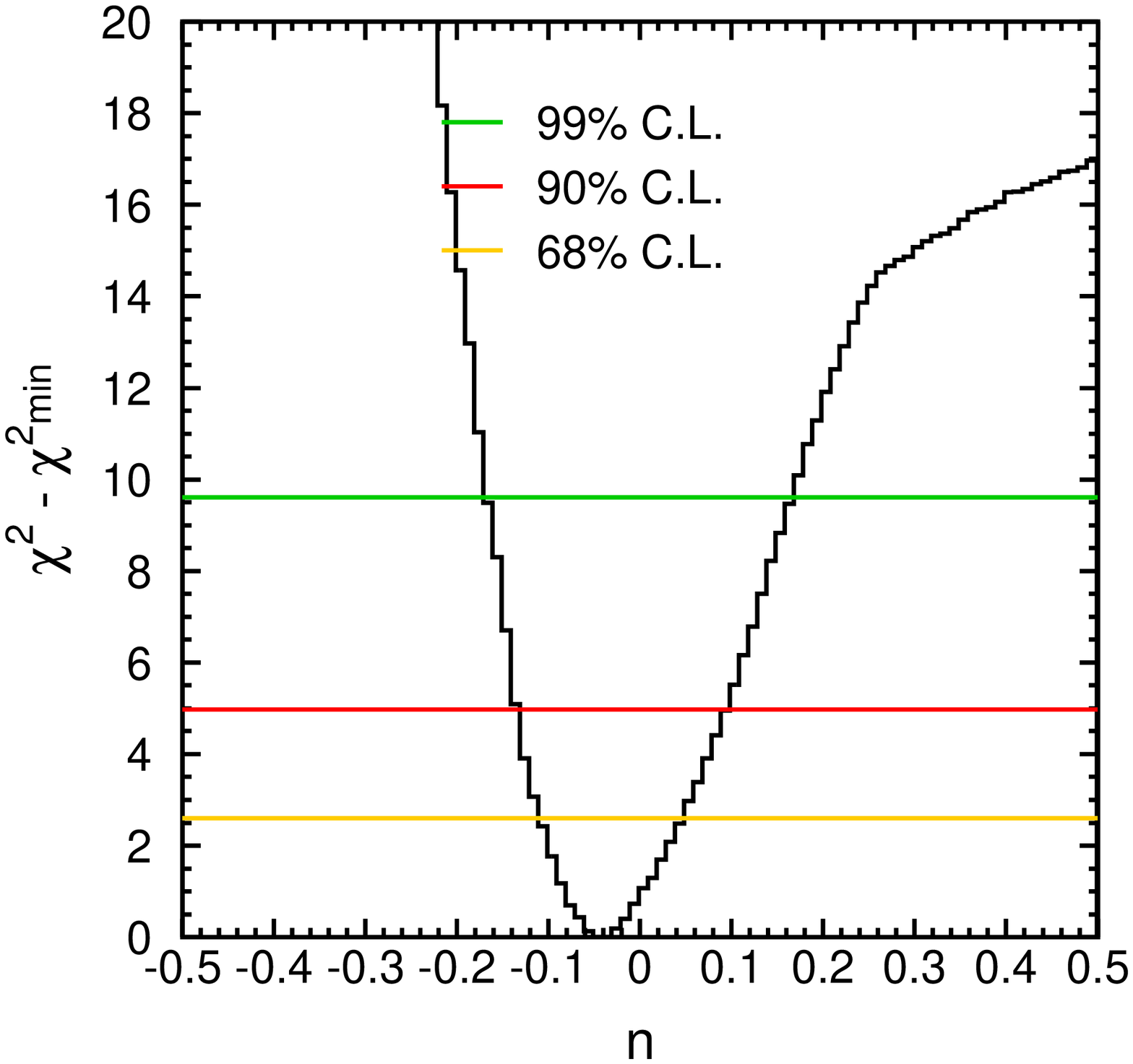}
 \caption{$\Delta m^2\rightarrow \Delta m^2\times 
\left(\frac{\rho_e}{\rho_o}\right)^n$ (including air path length). Upper plot shows 
relative-$\chi^2$ confidence level contours on the $\Delta m^2$ versus $n$ 
plane, obtained when taking into account both high and low density matter path 
lengths.  The lower plots display the $\chi^2-\chi^2_{min}$ contours, with confidence 
levels shown, at the best-fit parameter values. }
 \label{fig:powers_air}
\end{figure}

Finally, we again treat density dependence $n$ as a free parameter, this time taking 
air path length into account, once more assuming maximal mixing.  The results are 
shown in Figure~\ref{fig:powers_air}.  A minimum $\chi^2$ = 174.3/178 d.o.f. is found 
at ($n,\Delta m^2) = (-0.04, 1.95 \times 10^{-3} {\rm eV}^2)$. This result and the standard 
2-flavor oscillation result (i.e. with $n\equiv{0}$) do not differ significantly.

\section{Summary and Conclusions}
\label{sec:summary}
We used data from Super-Kamiokande-I to test mass-varying neutrino models, by considering oscillation effects where the effective mass of a neutrino is dependent on the matter environment of the neutrino flight path.  Our analysis assumed varying neutrino mass was due entirely to the electron density of the material in a neutrino's path, a possible consequence of the MaVaN theory.  Assuming a constant mixing angle in all environments, we substituted $\Delta m^2$ with $\Delta m^2_\times \left(\frac{\rho_e}{\rho_o}\right)^n$  (where $\rho_o=6.02\times10^{23}{\rm{e}}/{\rm{cm}}^3$) for specific values of $n$ suggested by theorists, as well as with $n$ as a fitted parameter. Two different models were considered.
In the first analysis, we neglected the portion of the neutrino flight path in low density matter (air), and assumed mass-varying effects occur only in high density matter (rock). For the second analysis, we allowed oscillation effects in all portions of the neutrino path, but (in contrast to the conventional oscillation analysis, where only the geometric path length is considered) took into account mass densities. The primary difference in the two hypothesis lies in the effective path lengths for near-horizontal downward-going neutrinos.  

In the ``pull" fitting procedure used, some of the systematic errors are allowed to 
vary in the fit. When fitted systematics are forced significantly outside a reasonable 
range of values, it is an indication that the model used is disfavored by the data. 
For the high-density-only oscillation model, some of the systematics, particularly the 
$\nu_\mu / \nu_e$ ratio for energies below 5 GeV, adjusted to values beyond a 
reasonable deviation from the expected value.  This does not happen when fitting to 
the conventional oscillation model, and results in a larger chi-squared value for the 
mass-dependent models.  Based on relative chi-squared values in our results, the 
hypothesis that neutrinos oscillate only in high density matter is disfavored relative 
to conventional oscillations by at least 3.5 $\sigma$, without requiring additional 
effects.

For the second analysis, where oscillations in low density matter were included, a 
mass difference with electron density dependence is disfavored at more than the 
3.8-$\sigma$ level when compared with neutrino oscillations with a fixed mass 
difference, for all fixed values of $n$ tested.  In addition, a freely varying density 
dependence analysis was performed, assuming maximal mixing.  It produced a minimum 
chi-squared value of $\chi^2_{min}$ = 174.3/178 d.o.f. at ($n = -0.04, \Delta m^2 = 
1.95 \times 10^{-3} eV^2)$, consistent (within 1.4 $\sigma$) with the $n=0$ (no 
density dependence) case.

The results of our analyses show no evidence that the environmental electron density influences the effective $\Delta m^2$ determined using Super-Kamiokande atmospheric neutrino data.  We have not explicitly considered models where the mixing angle is not constant in all densities, or which assume 3 flavor oscillations, nor do we exclude all variations of MaVaN models.  We find that conventional density independent \mutau oscillations are sufficient to explain the atmospheric neutrino data.

\section{Acknowledgements}

We gratefully acknowledge the cooperation of the Kamioka Mining and Smelting Company.  
The Super-Kamiokande experiment was built and has been operated with funding from the 
Japanese Ministry of Education, Science, Sports and Culture, and the United States 
Department of Energy.  We gratefully acknowledge individual support by the National 
Science Foundation, and the Polish Committee for Scientific Research.

\bibliography{superkmavans-revised}

\end{document}